\documentclass[10pt]{amsart}
\usepackage{epsfig}
\usepackage{amsmath,amssymb}
\numberwithin{equation}{section}

\newtheorem{theorem}{Theorem}[section]

\newtheorem{proposition}[theorem]{Proposition}

\allowdisplaybreaks \numberwithin{equation}{section}

\def\Im{\mathop{\rm Im}\nolimits}
\def\Re{\mathop{\rm Re}\nolimits}

\def\diag{\mathop{\rm Diag}\nolimits}

\def\im{\mathop{\imath}}
\def\adj{\mathop{\rm Adj}\nolimits}

\begin{document}
\title{Finite-gap integration of the  $SU(2)$ Bogomolny equations}
\author{H.W. Braden}
\address{School of Mathematics, Edinburgh University, Edinburgh.}
\email{hwb@ed.ac.uk}
\author{V.Z. Enolski}
\address{Department of Mathematics and Statistics, Concordia
University, Montreal.
\newline
On leave: Institute of Magnetism, National Academy of Sciences of
Ukraine.} \email{vze@ma.hw.ac.uk}

\begin{abstract}
The ADHMN construction of magnetic monopoles is given in terms of
the (normalizable) solutions of an associated Weyl equation. We
focus here on solving this equation directly by algebro-geometric
means. The (adjoint) Weyl equation is solved using an ans\"atz of
Nahm in terms of Baker-Akhiezer functions. The solution of Nahm's
equation is not directly used in our development.
\end{abstract}

\maketitle
\section{Introduction}
Consider the dimensional reduction to three dimensions of the four
dimensional Yang-Mills Lagrangian with gauge group $ SU(2)$ under
the assumption that all fields are independent of time. Upon
identifying the $a_4$-component of the gauge field with the Higgs
field $\Phi $ we obtain the three dimensional Yang-Mills-Higgs
Lagrangian
\[L= -\frac12 \mathrm{Tr}\, F_{ij} F^{ij}+\mathrm{Tr}\,D_{i}\Phi\, D^{i}\Phi.\]
Here $F_{ij}=\partial_i a_j -\partial_j a_i+[a_i,a_j]$ is the
curvature of the (spatial) connection of the gauge field
$a_i(\boldsymbol{x})$ and $D_i$  the covariant derivative
$D_i\Phi=\partial_i\Phi+[a_i,\Phi]$,
$\boldsymbol{x}=(x_1,x_2,x_3)\in \mathbb{R}^3$. We are interested
in configurations minimizing the energy of the system. These are
given by the {\it Bogomolny equation}
\begin{equation} D_i\Phi=\pm \sum_{j,k=1}^3\epsilon_{ijk}
F_{jk},\quad i=1,2,3\label{bogomolny}.\end{equation} A solution with the boundary
conditions
\[\left. \sqrt{-\frac12 \mathrm{Tr}\,\Phi(r)^2}\right|_{r\rightarrow\infty}\sim
1-\frac{n}{2r}+O(r^{-2}),\quad r=\sqrt{x_1^2+x_2^2+x_3^2}\] is called a {\it monopole} of  charge $n$.

In this note note we shall follow the
Atiyah-Drinfeld-Hitchin-Manin-Nahm (ADHMN) construction (see the
original papers \cite{nahm82},\cite{hitchin82} and the recent
review \cite{weinbyi06}). This has its origins in the construction
of instanton solutions to the (Euclidean) self-dual Yang-Mills
equations by Atiyah-Drinfeld-Hitchin-Manin (ADHM): here the
self-duality equations, partial differential equation in four
variables, are transformed to an algebraic matrix equation.
In the monopole setting the ADHMN construction reduces the
Bogomolny equation, again a partial differential equation but now
in three variables, to a system of ordinary differential
equations. Our interest here will be to integrate these by
algebro-geometric methods.

The standard approach to the integration of the Bogomolny equation
within the ADHMN construction consists of two stages. First, an
auxiliary equation known as Nahm's equation,
\begin{equation}
\frac{\mathrm{d}T_i(z)}{\mathrm{d}z}
=\frac12\sum_{j,k=1}^3\epsilon_{ijk}[T_j(z),T_k(z)], \quad
z\in[-1,1] \label{nahmeq}\end{equation} is integrated for $n\times n$ matrices $T_i(z)$
subject to
certain boundary conditions. The solution of this equation is then
used to define a differential operator
\begin{align}
\Delta^{\dagger}= \im 1_{2n}\frac{\mathrm{d}}{\mathrm{d}z}
-\sum_{j=1}^3( T_j(z)+\im x_j 1_n) \otimes \sigma_j,
\label{weylequ}\end{align} where $\sigma_j$ are the Pauli
matrices. The Higgs and gauge fields are then expressed  as
certain averages over the normalizable  solutions
$\boldsymbol{{v}}$ to the Weyl equation
$\Delta^{\dagger}\boldsymbol{{v}}(\boldsymbol{x},z)=0$. Nahm's
equation, introduced in the ADHMN construction, plays an important
role in many problems of mathematical physics and the integration
of this equation is of great significance. Its role in the
integration of the Bogomolny equation is nevertheless an auxiliary
one. In this paper we shall concentrate on solving of the Weyl
equation directly by algebro-geometric means using an ansatz
(again of Nahm) that has not been considered previously in this
light. Our new insight is that this ans\"atz  may be solved in
terms of a Baker-Akhiezer function. Although the associated
spectral problem is equivalent to that appearing in the
algebro-geometric integration of the Nahm equation we do not use
solutions of Nahm's equation directly in our development. To
achieve our result we implement the $\theta$-functional
integration of the Nahm equation by Ercolani and Sinha
\cite{ersi89} and our recent analysis \cite{bren06}. The
limitations of space in this volume prevent detailed examples
being given and a fuller exposition will be given elsewhere.

\section{The ADMHN construction}

Set
\begin{equation}
\Delta=\im \dfrac{d}{dz}+x-\im T_4+\boldsymbol{T}\cdot\boldsymbol{\sigma}
=\im \dfrac{d}{dz}-\im R,
\end{equation}
where
\begin{equation}
x=x_4+\im \boldsymbol{x}\cdot\boldsymbol{\sigma},\quad
T=T_4+\im \boldsymbol{T}\cdot\boldsymbol{\sigma},\quad
R=T+ix.
\end{equation}
We will often assume we have chosen a gauge such that $T_4=0$ and
that $x_4=0$. The ADMHN construction may be summarised in  the
following theorem.
\begin{theorem} [\bf ADMHN] The charge $n$
monopole solution of the Bogomolny equation is given by
\begin{align}
\Phi_{ab}(\boldsymbol{x})&=\im\int_{-1}^1dz\,z \boldsymbol{v}_a\sp\dagger(\boldsymbol{x},z)\boldsymbol{v}_b(\boldsymbol{x},z)
,\quad a,b=1,2,\label{higgs}\\
A_{i\, ab}(\boldsymbol{x})&=\int_{-1}^1dz\, \boldsymbol{v}_a\sp\dagger(\boldsymbol{x},z)
\frac{\partial}{\partial x_i} \boldsymbol{v}_b(\boldsymbol{x},z),
\quad
i=1,2,3,\quad a,b=1,2.\label{gauges}
\end{align}
Here the two ($a=1,2$) $2n$-column vectors\footnote{Throughout the
paper vectors are column-vectors and printed in bold, e.g.
$\boldsymbol{a}$; the superscript $\dagger$ means conjugated and
transposed, e.g for vector
$\boldsymbol{a}^{\dagger}=\overline{\boldsymbol{a}}^T$, and this
holds similarly for matrices. }
${{\boldsymbol{v}}}_{a}(\boldsymbol{x},z)
=({v}_1^{(a)}(\boldsymbol{x},z),\ldots,
{v}_{2n}^{(a)}(\boldsymbol{x},z))^T$ form an orthonormal basis on
the interval $z\in[-1,1]$
\begin{equation}\int_{-1}^1dz\, \boldsymbol{v}_a\sp\dagger(\boldsymbol{x},z)
\boldsymbol{v}_b(\boldsymbol{x},z)
=\delta_{ab}
,\label{norm}
\end{equation}
for the normalizable solutions to the the Weyl equation
\begin{align}\Delta\sp\dagger \boldsymbol{v}=0,
\end{align}
where $\Delta^{\dagger}$ is given by (\ref{weylequ}). The
normalizable solutions form a two-dimensional subspace of the
solution space
$(\boldsymbol{v}^{(1)}(\boldsymbol{x},z),\ldots,\boldsymbol{v}^{(2n)}(\boldsymbol{x},z))$.
The $n\times n$-matrices $T_j(z)$, called Nahm data,  satisfy
Nahm's equation (\ref{nahmeq}) and are required to satisfy the
following boundary conditions: they are regular at $z\in(-1,1)$;
have simple poles at $z=\pm1$, the residues of which form the
irreducible $n$-dimensional representation of the $su(2)$ algebra;
further
\begin{equation}
T_i(z)=-T_i^{\dagger}(z),\quad
T_i(z)=T_i^{T}(-z).\label{constraint}\end{equation}
\end{theorem}
A proof consisting of direct verification may be found for example
in the recent exposition by E.Weinberg and P.Yi \cite{weinbyi06}
and references therein.

The integrals in  (\ref{higgs}), (\ref{gauges}) and (\ref{norm})
may be computed in closed form \cite{panagopo83} in the following
way. Denote by
\begin{equation}
\mathcal{H}(\boldsymbol{x})=\boldsymbol{x}\cdot\boldsymbol{\sigma}\otimes
\mathrm{1}_n,\quad \mathcal{T}(z)=\imath
\boldsymbol{\sigma}\otimes\boldsymbol{T}(z),\label{pnotation}
\end{equation}
where $\boldsymbol{\sigma}=(\sigma_1,\sigma_2,\sigma_3)^T$,
$\boldsymbol{T}=(T_1,T_2,T_3)^T$ and
$\boldsymbol{x}\cdot\boldsymbol{\sigma}\otimes
\mathrm{1}_n=\sum_{i=1}\sp3x_i\sigma_i\otimes 1_n$,
$\boldsymbol{\sigma}\otimes\boldsymbol{T}(z)=\sum_{i=1}\sp3\sigma_i\otimes
T_i(z)$. Introduce the $2n\times 2n$ matrix
$\mathcal{Q}(z,\boldsymbol{x})$,
\begin{equation}
\mathcal{Q}(\boldsymbol{x},z)
=\frac{1}{r^2}\mathcal{H}(\boldsymbol{x})\mathcal{T}(z)\mathcal{H}(\boldsymbol{x})-\mathcal{T}(z).
\end{equation}
Then the following formulae of Panagopoulos (see appendix A) are
valid for the normalizable Weyl spinors,
$\boldsymbol{{v}}_{1,2}(\boldsymbol{x},z)$,
\begin{align}
&\int
{\boldsymbol{{v}}_{a}^\dagger}(\boldsymbol{x},z)
\boldsymbol{{v}}_{b}(\boldsymbol{x},z)\mathrm{d}z={\boldsymbol{{v}}_{a}^\dagger}
(\boldsymbol{x},z)
\mathcal{Q}^{-1}(\boldsymbol{x},z)\boldsymbol{{v}}_{b}(\boldsymbol{x},z);
\label{weylnorm}\\
& \int
z{\boldsymbol{{v}}_{a}^\dagger}(\boldsymbol{x},z)
\boldsymbol{{v}}_{b}(\boldsymbol{x},z)\mathrm{d}z={\boldsymbol{{v}}_{a}^\dagger}
(\boldsymbol{x},z)
\mathcal{Q}^{-1}(\boldsymbol{x},z)\left[z+2\mathcal{H}(\boldsymbol{x})\frac{\mathrm{d}}{
\mathrm{d}(r^2)}
 \right]\boldsymbol{{v}}_{b}(\boldsymbol{x},z);
\label{weylhiggs}\\
& \int
{\boldsymbol{{v}}_{a}^\dagger}(\boldsymbol{x},z)
\frac{\partial}{\partial x_i}\boldsymbol{{v}}_{b}(\boldsymbol{x},z)\mathrm{d}z \notag\\
&={\boldsymbol{{v}}_{a}^\dagger}(\boldsymbol{x},z)
\mathcal{Q}^{-1}(\boldsymbol{x},z)\left[\frac{\partial}{\partial
 x_i}+\mathcal{H}(\boldsymbol{x})\frac{zx_i+\imath (\boldsymbol{x}\times \nabla)_i }{ r^2}
 \right]\boldsymbol{{v}}_{b}(\boldsymbol{x},z).
\label{weylgauges}
\end{align}
Therefore only the boundary values of the normalized Weyl spinors
$\boldsymbol{{v}}_{1,2}(\boldsymbol{x},\pm1)$ together with their
derivatives need be computed to find solutions to the Bogomolny
equation. The Nahm data at these boundary values is also needed,
and this involves the residues noted above.

\section{The Nahm Ans\"atz}

Although we wish to solve
$$\Delta\sp\dagger \boldsymbol{v}=0$$ Nahm introduced an ans\"atz  that provides solutions to
$$\Delta \boldsymbol{w}=0$$ that we now recall.
Consider solutions of the form
\begin{equation}\label{Nansatz}
\boldsymbol{w}=(1_2+\boldsymbol{\hat u}(\boldsymbol{x})\cdot\boldsymbol{\sigma})\,e\sp{\im x_4z}|s>\otimes\,
\boldsymbol{\hat w(z)}
\end{equation}
where $|s>$ is an arbitrary normalized spinor not in $\ker
(1_2+\boldsymbol{\hat u}(\boldsymbol{x})\cdot\boldsymbol{\sigma})$
and $\boldsymbol{\hat u}(\boldsymbol{x})$ is (as we shall see) a
unit vector independent of $z$. Substituting in $\Delta
\boldsymbol{w}=0$ we find
$$0=|s>\otimes\left(\im \dfrac{d}{dz}+
\boldsymbol{\hat u}\cdot\boldsymbol{R}\right)\boldsymbol{\hat w(z)}+
\sigma_k|s>\otimes\left(\im {\hat u}\sp{k}\dfrac{d}{dz}+R\sp{k}+
\im(\boldsymbol{R}\times\boldsymbol{\hat u})\sp{k}\right)\boldsymbol{\hat w(z)}$$ and so we require
\begin{align}
0&=\left(\im \dfrac{d}{dz}+
\boldsymbol{\hat u}\cdot\boldsymbol{R}\right)\boldsymbol{\hat w(z)},\label{nahm1}\\
0&=\mathcal{L}_k\boldsymbol{\hat w(z)}:=\left(\im {\hat u}\sp{k}\dfrac{d}{dz}+R\sp{k}+
\im(\boldsymbol{R}\times\boldsymbol{\hat u})\sp{k}\right)\boldsymbol{\hat w(z)}.\label{nahm2a}
\end{align}
The consistency of these equations imposes various constraints.
First consider
\begin{align*}
\left[\mathcal{L}_1,\mathcal{L}_2
\right]&=(\im {\hat u}\sp1+{\hat u}\sp2{\hat u}\sp3)
\left(\dot{T_2}-[T_3,T_1]\right)-(\im {\hat u}\sp2-{\hat u}\sp1{\hat u}\sp3)
\left(\dot{T_1}-[T_2,T_3]\right)
\\
&\qquad-(1-({\hat u}\sp3)\sp2)\left(\dot{T_3}-[T_1,T_2]\right)+(1-\boldsymbol{\hat u}\cdot\boldsymbol{\hat u})\,\dot{T_3}.
\end{align*}
Thus provided $\boldsymbol{\hat u}(\boldsymbol{x})$ is a unit
vector and the $T_i$'s satisfy the Nahm equations we have
consistency of the equations $\mathcal{L}_k\boldsymbol{\hat
w(z)}=0$.

At this stage we introduce a convenient parameterization
(reflected in Hitchin's minitwistor construction). Let
$\boldsymbol{ y}\in \mathbb{C}\sp3$ be a null vector. We may
consider $\boldsymbol{ y}\in \mathbb{P}\sp2$ and parameterize
$\boldsymbol{ y}$ as
\begin{equation}\label{defhy} \boldsymbol{
y}=\left(\frac{1+\zeta^2}{2\im},\frac{1-\zeta^2}2,-\zeta \right).
\end{equation}
Then
$$\boldsymbol{ y}\cdot \boldsymbol{\overline
y}=\frac{(1+|\zeta|\sp2)\sp2}2,\qquad \boldsymbol{
y}\cdot \boldsymbol{ y}=0.$$
The signs here have been chosen so that
$$L(\zeta):=2\im \boldsymbol{ y}\cdot\boldsymbol{T}=
(T_1+\im T_2)-2\im T_3\,\zeta+(T_1-\im T_2)\,\zeta\sp2.$$ In due
course we will see this to be our Lax matrix. Set
\begin{equation}\label{defhu}
\boldsymbol{\hat
u}=\boldsymbol{\hat
u}(\zeta):=\im\,\frac{\boldsymbol{ y}\times \boldsymbol{\overline
y}}{\boldsymbol{ y}\cdot \boldsymbol{\overline
y}}=\frac{1}{1+|\zeta|^2}\left(\im(\zeta-\overline{\zeta}),\;(\zeta+\overline{\zeta}),
1-|\zeta|^2\right).
\end{equation}
Then
$$\boldsymbol{\hat
u}\times\boldsymbol{ y}=-\im\boldsymbol{ y},\qquad
\boldsymbol{\hat u}\times\boldsymbol{\overline
y}=\im\boldsymbol{\overline y}.$$ The three vectors
$\Re(\boldsymbol{ y}),\Im(\boldsymbol{ y})$ and $ \boldsymbol{\hat
u}$ form an orthogonal basis in $\mathbb{R}^3$ with
$|\boldsymbol{\hat u}|=1$, whence any $\boldsymbol{
v}\in\mathbb{R}^3$ may be written as
\begin{equation*}\label{basisuy}\boldsymbol{ v}=\boldsymbol{\hat
u}\,(\boldsymbol{\hat u}\cdot \boldsymbol{v})+
\boldsymbol{\overline y}\left(\frac{\boldsymbol{ y}\cdot
\boldsymbol{v}}{ \boldsymbol{ y}\cdot\boldsymbol{\overline
y}}\right) + \boldsymbol{ y}\left(\frac{\boldsymbol{\overline
y}\cdot \boldsymbol{v}}{ \boldsymbol{ y}\cdot\boldsymbol{\overline
y}}\right).
\end{equation*}
In particular,
\begin{equation}\label{basisucy}\boldsymbol{ v}+\im
\boldsymbol{ v}\times\boldsymbol{\hat u}=\boldsymbol{\hat
u}\,(\boldsymbol{\hat u}\cdot \boldsymbol{v})+2
\boldsymbol{\overline y}\left(\frac{\boldsymbol{ y}\cdot
\boldsymbol{v}}{ \boldsymbol{ y}\cdot\boldsymbol{\overline
y}}\right) .
\end{equation}
 We record that
\begin{align*}
\overline{\boldsymbol{ y}(\zeta)}&=-{\overline\zeta}\sp2\, \boldsymbol{ y}(-1/{\overline
\zeta}),\qquad\boldsymbol{\hat u}(-1/{\overline
\zeta})=-\boldsymbol{\hat u}(\zeta),\\
\boldsymbol{\hat u}&=(-\im\zeta\sp{-1},\zeta\sp{-1},-1)-\frac{2\boldsymbol{ y}}
{\zeta(1+|\zeta|\sp2)}
=(\im\zeta,\zeta,1)+\frac{2{\overline\zeta}\boldsymbol{ y}}
{1+|\zeta|\sp2},\\
\boldsymbol{\hat u}\cdot\boldsymbol{T}&=
-\im\left[(T_1+\im T_2)\zeta\sp{-1}-\im T_3\right]-
\frac{2\boldsymbol{ y}\cdot\boldsymbol{ T}}
{\zeta(1+|\zeta|\sp2)}=
\im\left[(T_1-\im T_2)\zeta-\im T_3\right]-
\frac{2{\overline\zeta}\boldsymbol{ y}\cdot\boldsymbol{ T}}
{1+|\zeta|\sp2}.
\end{align*}

Parameterizing $\boldsymbol{\hat u}$ as above and using
(\ref{basisucy}) we may write
\begin{equation*}
\im \boldsymbol{\hat u}\dfrac{d}{dz}+\boldsymbol{R}+
\im\boldsymbol{R}\times\boldsymbol{\hat u}=\im \boldsymbol{\hat u}\dfrac{d}{dz}+
\boldsymbol{\hat
u}\,(\boldsymbol{\hat u}\cdot \boldsymbol{R})+2
\boldsymbol{\overline y}\left(\frac{\boldsymbol{ y}\cdot
\boldsymbol{R}}{ \boldsymbol{ y}\cdot\boldsymbol{\overline
y}}\right) =\boldsymbol{\hat u}\left(\im \dfrac{d}{dz}+
\boldsymbol{\hat u}\cdot\boldsymbol{R}\right)+2
\boldsymbol{\overline y}\left(\frac{\boldsymbol{ y}\cdot
\boldsymbol{R}}{ \boldsymbol{ y}\cdot\boldsymbol{\overline
y}}\right)
\end{equation*}
and as a consequence (\ref{nahm1}, \ref{nahm2a}) are equivalent to
\begin{align}
0&=\left(\im \dfrac{d}{dz}+
\boldsymbol{\hat u}\cdot\boldsymbol{R}\right)\boldsymbol{\hat w(z)},\label{nahm1b}\\
0&=\left(\boldsymbol{ y}\cdot
\boldsymbol{R}\right)\boldsymbol{\hat w(z)}.\label{nahm2b}
\end{align}
The remaining consistency to be checked is then
\begin{align*}
\left[\im \dfrac{d}{dz}+
\boldsymbol{\hat u}\cdot\boldsymbol{R},\,
\boldsymbol{ y}\cdot
\boldsymbol{R}
\right]
&=
\im {\boldsymbol{y}\cdot{\dot{\boldsymbol{T}}}}
+
\left[\boldsymbol{\hat u}\cdot\boldsymbol{T},\,
{\boldsymbol{y}\cdot{\boldsymbol{T}}}
\right]
=0.
\end{align*}
which upon use of $\boldsymbol{\hat u}\times\boldsymbol{
y}=-\im\boldsymbol{ y}$ is equivalent to Nahm's equations.

Equally from
$$\boldsymbol{\hat u}\cdot\boldsymbol{R}=
-\im\left[(R_1+\im R_2)\zeta\sp{-1}-\im R_3\right]-
\frac{2\boldsymbol{ y}\cdot\boldsymbol{ R}}
{\zeta(1+|\zeta|\sp2)}=
\im\left[(R_1-\im R_2)\zeta-\im R_3\right]-
\frac{2{\overline\zeta}\boldsymbol{ y}\cdot\boldsymbol{ R}}
{1+|\zeta|\sp2}$$ we may write the equations as
\begin{align*}
0&=\left( \dfrac{d}{dz}
+\left[(R_1-\im R_2)\zeta-\im R_3\right]\right)\boldsymbol{\hat w(z)}
=\left( \dfrac{d}{dz}+M
+\im\left[(x_1-\im x_2)\zeta-\im x_3\right]\right)\boldsymbol{\hat w(z)},\\
0&=\left(\boldsymbol{ y}\cdot
\boldsymbol{R}\right)\boldsymbol{\hat w(z)},
\end{align*}
{where} \begin{equation} M=(T_1-\im T_2)\zeta-\im T_3.
\end{equation}

The equations we have obtained are just the Lax equations
\begin{align*}
0&={2\im}\left(\boldsymbol{ y}\cdot
\boldsymbol{R}\right)\boldsymbol{\hat w(z)}
=\left(L(\zeta)-\eta\right)\boldsymbol{\hat w(z)},
\qquad
\eta={2\boldsymbol{y}\cdot{\boldsymbol{x}}},\\
0&=\left(\im \dfrac{d}{dz}+
\boldsymbol{\hat u}\cdot\boldsymbol{R}\right)\boldsymbol{\hat w(z)},
\intertext{and}
\dot L&=[L,M].
\end{align*}
From the first of these we see that
$$0=\det\left(L(\zeta)-\eta\right),$$
which gives the equation of the spectral curve $\mathcal{C}$. Upon
using $\overline{\boldsymbol{ y}(\zeta)}=-{\overline\zeta}\sp2\,
\boldsymbol{ y}(-1/{\overline \zeta})$ we see from
$$0=\det\left(L(\zeta)-\eta\right)\sp\dagger
=\det\left(L(\zeta)\sp\dagger-\overline{\eta}\right)=
\det\left({2\im\overline{\boldsymbol{ y}(\zeta)}\cdot
\boldsymbol{T}}-\overline{\eta}\right)=
\det\left({-2\im{\overline\zeta}\sp2{\boldsymbol{ y}(-1/{\overline
\zeta})}\cdot
\boldsymbol{T}}-\overline{\eta}\right)
$$
that the spectral curve is invariant under
$$(\zeta,\eta)\rightarrow
(-\frac1{\overline\zeta},-\frac{\overline\eta}{{\overline\zeta}\sp2}).$$
The spectral curve then has the form
\begin{equation}
\eta^n+a_1(\zeta)\eta^{n-1}+\ldots+a_n(\zeta)=0, \quad
\mathrm{deg}\, a_k(\zeta) \leq 2k\label{curve1},
\end{equation}
and the genus of $\mathcal{C}$ is $g=(n-1)^2$.

It is worth remarking that Nahm's ans\"atz only yields solutions
of $\Delta \boldsymbol{w}=0$ and does not yield solutions of
$\Delta\sp\dagger \boldsymbol{v}=0$.

\subsection{Strategy of Solution}

The strategy for constructing solutions involves 3 steps. We have
seen that finding solutions to $\Delta \boldsymbol{w}=0$ reduces
to solving
\begin{align}
0&=\left(L(\zeta)-\eta\right)\boldsymbol{\hat w(z)},\label{eqlax1}\\
0&=\left( \dfrac{d}{dz}+M
\right)\boldsymbol{\hat w(z)},\label{eqlax2}
\end{align}
upon using the (slightly modified) ans\"atz
\begin{equation*}
\boldsymbol{w}=(1_2+\boldsymbol{\hat u}(\boldsymbol{x})\cdot\boldsymbol{\sigma})\,
e\sp{-\im z\left[(x_1-\im x_2)\zeta-\im x_3 -x_4\right]}|s>\otimes\,
\boldsymbol{\hat w(z)}.
\end{equation*}
Here  $\boldsymbol{\hat u}(\boldsymbol{x})$ is a unit vector and
$\eta={2\boldsymbol{y}\cdot{\boldsymbol{x}}}$. We might construct
a solution as follows.

\begin{enumerate}
\item Given a spectral curve
    $0=\det\left(L(\zeta)-\eta\right)$ and a position
    $\boldsymbol{x}$ we substitute
    $\eta={2\boldsymbol{y}\cdot{\boldsymbol{x}}}$ using the
    expression for $\boldsymbol{y}$ in terms of $\zeta$. This
    is an equation of degree $2n$ in $\zeta$ which we shall
    refer to as the \emph{Atiyah-Ward} constraint, this
    equation having appeared in their work. The $2n$ solutions
    give us $2n$ associated values $\boldsymbol{\hat
    u}\sp{a}$, $a=1,\dots,2n$. For each of these we solve for
    $\boldsymbol{\hat w(z)}$ yielding a $2n\times 1$ matrix
    $\boldsymbol{w}\sp{a}$. Taking each of the $2n$ solutions
    we obtain a $2n\times 2n$ matrix of solutions $W$.

\item As $0=\Delta W=\im\left(\frac{d}{dz}-R\right)W$, then
\begin{align*}
\frac{d}{dz}W&=RW,\qquad
\frac{d}{dz}W\sp\dagger=W\sp\dagger R,\qquad
\frac{d}{dz}\left(W\sp\dagger\right)\sp{-1}
=-R\left(W\sp\dagger\right)\sp{-1},
\end{align*}
whence
\begin{equation}
0=\Delta \sp\dagger\left(W\sp\dagger\right)\sp{-1}=
\im\left(\frac{d}{dz}+R\right)\left(W\sp\dagger\right)\sp{-1}.
\end{equation}
So given $W$ we may construct
$V=\left(W\sp\dagger\right)\sp{-1}$.

\item To reconstruct the gauge and Higgs fields using the
    formulae of the previous section we must extract from $V$
    the two normalizable solutions.

\end{enumerate}

The new insight that the study of integrable systems brings to
this problem is that $\boldsymbol{\hat w(z)}$ may be understood as
a Baker-Akhiezer function and constructed explicitly. Before
considering this, we conclude the section by noting Nahm's
construction for $\boldsymbol{\hat w(z)}$.

\subsection{Constructing $\boldsymbol{\hat w(z)} $ using the adjoint equation }

We begin with several simple observations. First, assuming $L$ is
invertible, the Lax equation $\dot L=[L,M]$ means also that
\begin{align*}
\frac{d}{dz} L\sp{-1}&=-L\sp{-1}\dot L L\sp{-1}=[L\sp{-1},M],\\
\frac{d}{dz} \adj L&=\frac{d}{dz}\left(\det(L)\,L\sp{-1}\right) =[
\adj L,M].
\end{align*}
Second, suppose $\lambda_i$ is an eigenvalue of $L$ with
    associated eigenvector $f_i$, $Lf_i=\lambda_i f_i$. Then
    $f_i$ is only determined up to a scale $f_i\rightarrow f_i
    h_i(z)$ which may differ from eigenvector to eigenvector.
    Set
$$F=(f_1,\dots,f_n),\qquad
\Lambda=\diag(\lambda_1,\ldots,\lambda_n).$$ Then
\begin{equation*}LF=F\Lambda \label{eqLF}
\end{equation*}
is compatible with the Lax equation if and only if $F=F(z)$ is
governed by
\begin{equation}\label{eqldta}
\left(\frac{d}{dt}+M\right)F=F\diag(\alpha_1,\ldots,\alpha_n),
\end{equation}
for some $\alpha_i(z)$. Conversely, given a solution of this
equation we may reconstruct $L$ satisfying $\dot L=[L,M]$ via
$L=F\Lambda F\sp{-1}$. Third, if $\lambda_i$ is an eigenvalue of
$L$ we may construct a corresponding eigenvectors $f_i$ via
\begin{equation}
\label{nevw}
f_i=\adj(L-\lambda_i)\nu h_i(z),
\end{equation} where $\nu$ is any
constant vector. This follows as
$$(L-\lambda_i )f_i=(L-\lambda_i
)\adj(L-\lambda_i)\nu h_i(z)=\det(L-\lambda_i)\nu h_i(z)=0.$$ With
such eigenvectors $f_i$  we see that \begin{align*} \dot
F&=[\adj(L-\lambda_i),M]\nu h_i(z)+\adj(L-\lambda_i)\nu \dot
h_i(z)\\&=-MF+F h_i\sp{-1}\dot h_i+\adj(L-\lambda_i)M\nu
h_i(z)\end{align*} and for this to be of the form (\ref{eqldta})
we require
$$0=\adj(L-\lambda_i)\nu(\dot h_i-h_i\alpha_i)+\adj(L-\lambda_i)M\nu
h_i(t).$$ Taking the inner product with an arbitrary vector
$\mu$ then yields the differential equation
$$h_i\sp{-1}\frac{dh_i}{dz}=\alpha_i(z)-
\frac{\mu\sp{T}\adj(L-\lambda_i)M\nu}{\mu\sp{T}\adj(L-\lambda_i)\nu}.$$
Therefore requiring the differential equation for $F$ leads to a
differential equation for $h_i$. Suppose we write
$$h_i(z)=\frac{\exp\left[-\theta_i(z)+\int\sp{z}\alpha_i(z)\,dz
\right]}{\sqrt{\mu\sp{T}\adj(L-\lambda_i)\nu}}
$$
then
$$h_i\sp{-1}\frac{dh_i}{dz}=\alpha_i(z)-
\frac{d\theta_i}{dz}-\frac12
\frac{\mu\sp{T}\left[\adj(L-\lambda_i),M\right]\nu}{\mu\sp{T}\adj(L-\lambda_i)\nu}$$
which provides a solution if $$ \frac{d\theta_i}{dz}=\frac12
\frac{\mu\sp{T}\left\{M,\adj(L-\lambda_i)\right\}\nu}{\mu\sp{T}\adj(L-\lambda_i)\nu}
.$$

Nahm's approach to construct $\boldsymbol{\hat w(z)} $ was to
express this in the form (\ref{nevw}) together with the one
dimensional differential equations for $\theta_i$. This method has
only been implemented  in the charge $2$ case and we now propose
an alternative approach.

\section{A Spectral Problem}
We have identified $\boldsymbol{\hat w(z)} $ with the
Baker-Akhiezer function and now must ask whether this can be
constructed. Set
\begin{align*}
A_{-1}&=T_1+iT_2,\
  A_0   =-2iT_3,\
  A_{1} =T_1-iT_2,\\
\intertext{and so}
  L(\zeta)&=A_{-1}+A_0\zeta+A_1\zeta\sp2,\qquad
  M=\frac{1}{2}A_0+A_1\zeta.
\end{align*} Viewing equation (\ref{eqlax2}) as a spectral problem
$$
\left( \dfrac{d}{dz}+\frac12 A_0(z)
\right)\boldsymbol{\hat w(z)}=-\zeta A_1(z)\boldsymbol{\hat w(z)},
$$
we seek to solve this. The $z$-dependence of the right hand side
means however this is not a standard eigenvalue problem, but it
may be reduced to such using a trick of \cite{ersi89}. With the
notation introduced, Nahm's equation yield
\begin{equation}\frac{{d}}{{d}
z}A_1(z)=\frac12[A_0(z),A_1(z)]\end{equation} and so by
introducing the matrix $C(z)$ with \begin{equation*}
\frac{{d}}{{d} z}C(z)=\frac12 A_0(z)C(z), \qquad
C(0)=1_n\end{equation*}
we may write
$$A_1(z)=C(z)A_1(0)C(z)\sp{-1}.$$
Then upon performing a gauge transformation
$$Q_0(z)=C(z)A_0(z)C(z)\sp{-1},\qquad
\Phi(z)=C(z)\sp{-1}\hat w(z)$$ we obtain the spectral problem
\begin{align} \label{standard}
\left( \dfrac{d}{dz}+Q_0(z)
\right)\boldsymbol{\Phi}(z)&=-\zeta A_1(0)\boldsymbol{\Phi}(z).
\end{align}
Here
$\boldsymbol{\Phi}(z)=\boldsymbol{\Phi}(\zeta,\eta,z)=\boldsymbol{\Phi}(P,z)$
is given by the Baker-Akhiezer function on the curve,
$P=(\zeta,\eta)\in\mathcal{C}$. Then
$$\boldsymbol{\hat w(z)}=C(z)\boldsymbol{\Phi}(\zeta,\eta,z)$$
and
\begin{align*}
\boldsymbol{w}&=(1_2+\boldsymbol{\hat u}(\boldsymbol{x})\cdot\boldsymbol{\sigma})\,
e\sp{-\im z\left[(x_1-\im x_2)\zeta-\im x_3 -x_4\right]}|s>\otimes\,
C(z)\boldsymbol{\Phi}(\zeta,\eta,z),\\
&=1_2\otimes C(z) \left(
(1_2+\boldsymbol{\hat u}(\boldsymbol{x})\cdot\boldsymbol{\sigma})\,
e\sp{-\im z\left[(x_1-\im x_2)\zeta-\im x_3 -x_4\right]}|s>\otimes\,
\boldsymbol{\Phi}(\zeta,\eta
=\frac{2\boldsymbol{y}\cdot{\boldsymbol{x}}}{\zeta},z)
\right)
\end{align*}
Again, if we group all $2n$ solutions $\boldsymbol{\Phi}$ together
into a $n\times 2n$ matrix $\boldsymbol{\hat\Phi}$ we then obtain
$$W=\left(1_2\otimes C(z)\right) \varphi,
\qquad \varphi=(1_2+\boldsymbol{\hat u}(\boldsymbol{x})\cdot\boldsymbol{\sigma})\,
e\sp{-\im z\left[(x_1-\im x_2)\zeta-\im x_3 -x_4\right]}|s>\otimes \boldsymbol{\hat \Phi}
$$
where $\varphi$ is a $2n\times 2n$ matrix. Then
$$V\sp\dagger =W\sp{-1}=\varphi\sp{-1} (1_2\otimes C(z)\sp{-1})$$
will be in terms of the Baker-Akhiezer function. It remains then
to construct $\boldsymbol{\Phi}$.

\subsection{The Baker-Akhiezer function}
By a constant gauge transformation we may assume that $A_1(0)$ is
diagonal. Its behaviour may be read from the spectral curve
(\ref{curve1}),
$$A_1(0)=\mathrm{Diag}\left(\rho_1,\ldots,
\rho_m\right),\quad
\rho_m=\mathrm{Res}_{P\rightarrow\infty_m}\frac{\eta}{\zeta},
$$
where $\infty_m$ ($m=1,\ldots,n$) are the $n$ points above
$\zeta=\infty$. Thus the integration of the Adjoint Weyl equation
reduces to the matrix spectral problem (\ref{standard}). The same
problem appeared in \cite{ersi89} and \cite{bren06} when focussing
on the algebro-geometric integration of the Nahm equation and we
shall use the results of our recent paper \cite{bren06} for the
integration of the Weyl equation.

Let $\theta$ be the canonical $\theta$-function of the curve
$\mathcal{C}$ and let $\tau$ be its period matrix. The period
lattice is then generated by $\Lambda=(1_g,\tau)$ and
\[ \theta(\boldsymbol{w})=\sum_{\boldsymbol{k}\in\mathbb{Z}^g} \mathrm{exp}\left\{
\imath\pi\boldsymbol{k}^T \tau \boldsymbol{k}+2\imath\pi
\boldsymbol{w}^T\boldsymbol{k} \right\}.
\]
Denote by $\Theta=\{\boldsymbol{w}|\theta(\boldsymbol{w})=0\}$ the
$\theta$-divisor in the Jacobi variety of the curve $\mathcal{C}$,
$\mathbb{C}^g/\Lambda$.

\begin{theorem} Let $\boldsymbol{\Phi}(P,z)=(\Phi_1(P,z),\ldots,
\Phi_n(P,z))^T$ be the eigenfunction (or Baker-Akhiezer function)
of the standard spectral problem (\ref{standard}). The components
$\Phi_j(P,z)$  are given by
\begin{equation}\label{newbafnch}
\Phi _{j }\left( P,z\right) =g_{j }(P)\, \frac{ \theta \left(
\boldsymbol{\phi} (P)-\boldsymbol{\phi}(\infty_{j
})+(z+1)\,\boldsymbol{U}-\widetilde{\boldsymbol{K}}\right) \theta
\left(-\widetilde{\boldsymbol{K}} \right) } { \theta
\left(\boldsymbol{\phi} (P)-\boldsymbol{\phi}(\infty_{j })
-\widetilde{\boldsymbol{K}}\right) \theta \left(
(z+1)\,\boldsymbol{U}-\widetilde{\boldsymbol{K}}\right) }\,
e^{z\,\int\limits_{P_0}^{P}\gamma_{\infty}-z\,\nu_j}.
\end{equation}
Here $\boldsymbol{\phi}(P)$ is the Abel map, $z\in(-1,1)$, and
$P\in\mathcal{C}$. The vector $\widetilde{\boldsymbol{K}}$ is
defined by
$$\widetilde{\boldsymbol{K}}=
\boldsymbol{K}+\boldsymbol{\phi}\left((n-2)
\sum_{k=1}\sp{n}\infty_k\right),$$ where $\boldsymbol{K}$ is the
vector of Riemann constants. We have that
\begin{enumerate}\item $\widetilde{\boldsymbol{K}}$ is independent of the choice
of base point of the Abel map; \item $
\theta(\widetilde{\boldsymbol{K}})=0$;
\item $2\widetilde{\boldsymbol{K}} \in \Lambda$; \item for $n\ge3$ we
have $ \widetilde{\boldsymbol{K}}\in \Theta_{\rm singular}$.
\end{enumerate}
For each $j$ the function $g_j(P)$ is meromorphic on
$\mathcal{C}$, $g_j(\infty_j)=1$, and has a zero-divisor of degree
$g+n-1$ that includes the $n-1$ points
$(\infty_1,\ldots,\widehat{\infty_j},\ldots,\infty_n)$.

The matrix $Q_0(z)$ (which has poles of first order at $z=\pm1$)
is given by
\begin{equation}\label{ourq0}
Q_{0}(z)_{jl}  = \epsilon_{jl}\,\frac{\rho_{j}-
\rho_{l}}{\mathcal{E}(\infty_j,\infty_l)}\,e\sp{i\pi\boldsymbol{\tilde
q}\cdot(\boldsymbol{\phi}(\infty_l)-\boldsymbol{\phi}(\infty_j))}\,
    \,\frac{\theta(\boldsymbol{\phi}(\infty_{l}) -\boldsymbol{\phi}(
    \infty_{j}) + (z+1)\boldsymbol{U} - \widetilde{\boldsymbol{K}})}{\theta(
    (z+1)\boldsymbol{U}
    - \widetilde{\boldsymbol{K}})}\,e^{z(\nu_{l} - \nu_{j})}.
\end{equation}
Here $E(P,Q)=\mathcal{E}(P,Q)/\sqrt{\mathrm{d}x(P)\mathrm{d}x(Q)}$
is the Schottky-Klein prime form, $\boldsymbol{U} -
\widetilde{\boldsymbol{K}}=\frac12\boldsymbol{\tilde
p}+\frac12\tau\boldsymbol{\tilde q}$ ($\boldsymbol{\tilde p}$,
$\boldsymbol{\tilde q}\in\mathbb{Z}\sp{g}$) is a non-singular even
$\theta$-characteristic, and $\epsilon_{jl}=\epsilon_{lj}=\pm1$ is
determined (for $j<l$)
 by
$\epsilon_{jl}=\epsilon_{jj+1}\epsilon_{j+1j+2}\dots\epsilon_{l-1l}$.
The $n-1$ signs $\epsilon_{jj+1}=\pm1$ are arbitrary.
\end{theorem}
In passing we note that a formula with similar features was
obtained by Dubrovin \cite{dubrovin77} when giving a
$\theta$-functional solution to the Euler equation describing
motion of the $n$-dimensional rigid body. The essential difference
is that the curve $\mathcal{C}$ here should be subjected to the
the following three constraints
$\mathbf{H1},\mathbf{H2},\mathbf{H3}$ of Hitchin who showed a
bijection between such curves and magnetic monopoles
\cite{hitchin83}:

$\mathbf{H1}$ $\mathcal{C}$ admits the involution:
$(\zeta,\eta)\longrightarrow(-1/\overline{\zeta},-\overline{\eta}/\overline{\zeta}^2)$.

$\mathbf{H2}$ Let $\gamma_{\infty}(P)$ be the unique differential
 of the second kind on $\mathcal{C}$ defined by the conditions
\[ \gamma_{\infty}(P)\vert_{P\to\infty_i}=\left(\frac{\rho_i}{\xi^2}+O(1)\right)\mathrm{d}\xi,\quad
\oint_{\mathfrak{a}_k}\gamma_{\infty}(P)=0,\quad i, k=1,\ldots,g,
\] where $\xi$ is a local coordinate and
$\rho_i=\mathrm{Res}_{P\rightarrow\infty_i} \eta/\zeta$. Then
$\mathfrak{b}$-periods defining the winding vector
$\boldsymbol{U}$ are to be half-periods,
\begin{equation}
\boldsymbol{U}=\frac{1}{2 \pi\imath} \left(
\oint_{\mathfrak{b}_1}\gamma_{\infty},\ldots,
\oint_{\mathfrak{b}_g}\gamma_{\infty}
\right)^T=\frac12\boldsymbol{n}+\frac12\tau\boldsymbol{m}.
\end{equation}
The vectors $\boldsymbol{n}, \boldsymbol{m}\in \mathbb{Z}^g$ are
called  {\it Ercolani-Sinha vectors}. They should be {\it
primitive}, i.e. $s\boldsymbol{U}$ belongs to the period lattice
$\Lambda$ if and only if $s=0$ or $s=2$ (equivalently,
$z=s-1=\pm1$). (Hitchin's original constraint was  reformulated to
this form in \cite{bren06}.)

$\mathbf{H3}$ All components of the Baker-Akhiezer function
$\Phi_j(P,z)$ are real and smooth for $z\in (-1,1)$.

Bringing the previous results together then yields:
\begin{proposition}
Let $\boldsymbol{w}^{(k)}(\boldsymbol{x},z)$, $k=1,\ldots,2n$ be
the column vectors
\begin{equation}
\boldsymbol{w}^{(k)}(\boldsymbol{x},z)=
(1_2+\boldsymbol{\hat u}(P_k)\cdot\boldsymbol{\sigma})\,
e\sp{-\im z\left[(x_1-\im x_2)\zeta-\im x_3 -x_4\right]}|s>\otimes\,
C(z)\boldsymbol{\Phi}(P_k,z)
\end{equation}
where $P_k=(\zeta_k,\eta_k)\in\mathcal{C}$ are solutions to the
Atiyah-Ward constraint, $C(z)^{-1}$ is the  fundamental solution
to the ODE
\[\frac{\mathrm{d}}{\mathrm{d} z} C(z)^{-1}+\frac12 Q_0(z) C(z)^{-1}=0 \]
normalized by the condition $C(0)=1_n$, and the $n\times n$-matrix
$Q_0(z)$ and $n$-vector $\boldsymbol{\Phi}(P,z)$ are given by the
$\theta$-functional formulae (\ref{ourq0}) and (\ref{newbafnch})
respectively. Then the $2n\times 2n$ matrix
\begin{equation}V(\boldsymbol{x},z)=
\left\{\left(\boldsymbol{w}^{(1)}(\boldsymbol{x},z),\boldsymbol{w}^{(2)}(\boldsymbol{x},z),\ldots,
\boldsymbol{w}^{(2n)}(\boldsymbol{x},z)\right)^{-1}\right\}^{\dagger}
\end{equation}
defines the fundamental solution to the Weyl equation,
$\Delta^{\dagger}V=0$.

\end{proposition}

\section{Conclusions}

Although nonabelian magnetic monopoles have been objects of
fascination for some decades now, very few explicit solutions are
known. This note fits into our longer programme of seeing how far
the techniques from integrable systems will allow us to construct
such solutions. Here we have considered the explicit construction
of magnetic monopoles using algebro-geometric constructions coming
from integrable systems. Previous studies along these lines have
focussed on the construction of solutions to Nahm's equation which
is an auxiliary problem to that of the explicit integration of the
Bogomolny equations. Although the ADHM construction is based upon
normalizable solutions of the equation
$\Delta^{\dagger}\boldsymbol{v}=0$ an ans\"atz of Nahm naturally
gives solutions of the adjoint equation $\Delta\boldsymbol{w}=0$:
the matrices of fundamental solutions of these equations are
related by $V=\left(W\sp\dagger\right)\sp{-1}$. Here we have
expressed $\boldsymbol{w}$ in terms of a Baker-Akhiezer function
and given explicit expressions for this. Assuming one has a
spectral curve these expressions may be evaluated algorithmically.
Unfortunately the curves characterizing magnetic monopoles are
often restricted by transcendental constraints (\textbf{H2},
\textbf{H3} above), but this is a separate and interesting story
to the one presented here. Finally we have not addressed here the
remaining problem of extracting the normalizable solutions from
this data. Details and examples of this approach will be given
elsewhere.

\section{Acknowlegments} Both authors wish to acknowledge the partial
support of the European Science Foundation Programme MISGAM
(Methods of Integrable System, Geometry and Applied Mathematics)
and V.Z.E further thanks the ENIGMA network for additional support
while at Imperial College.

\appendix

\section{The Panagopolous formulae}

We have
$$
\Delta=\im \dfrac{d}{dz}+x_4+\im\boldsymbol{x}\cdot\boldsymbol{\sigma} -\im
T_4+\boldsymbol{T}\cdot\boldsymbol{\sigma}.$$ Thus
$$\Delta\sp\dagger=
\im
\dfrac{d}{dz}+x_4-\im\boldsymbol{x}\cdot\boldsymbol{\sigma}-\im
T_4-\boldsymbol{T}\cdot\boldsymbol{\sigma}.$$ Set
\[ \Delta^{\dagger}=\im\left[1_{2n}\frac{d}{dz}-\im x_4-T_4+\mathcal{H}+\mathcal{F}\right]\]
with Hermitian
\[  \mathcal{H}=-\sum_{j=1}^3 x_j \sigma_j\otimes 1_n,\qquad  \mathcal{F}=\imath \sum_{j=1}^3 \sigma_j\otimes T_j .\]
Then if $\boldsymbol{v}$ is any solution of
$\Delta^{\dagger}\boldsymbol{v}=0$ we have that
\begin{align*}
1_{2n}\frac{d}{dz}\boldsymbol{v}=\left[\im x_4+T_4-(\mathcal{H}+\mathcal{F})\right]\boldsymbol{v}.
\end{align*}

For completeness we prove here the Panagopolous formulae
\cite{panagopo83} using the method described in that reference
(extending very slightly to the case $x_4,T_4$ possibly nonzero).
These integral formulae reduce to the problem of finding for any
given operator $\mathcal{A}$ and any two solutions
$\boldsymbol{{v}}_{a,b}$ of $\Delta^{\dagger}\boldsymbol{v}=0$ an
operator $\mathcal{B}$ such that
\begin{equation}
\boldsymbol{{v}}_a\sp\dagger
\mathcal{A}\boldsymbol{{v}}_b=\frac{d}{dz} \left(\boldsymbol{{v}}_a\sp\dagger
\mathcal{B}\boldsymbol{{v}}_b\right).
\end{equation}
In this case
\begin{align*}
\boldsymbol{{v}}_a\sp\dagger\mathcal{A}\boldsymbol{{v}}_b
&=\frac{d \boldsymbol{{v}}_a\sp\dagger}{dz}
\mathcal{B}\boldsymbol{{v}}_b
+\boldsymbol{{v}}_a\sp\dagger
\frac{d\mathcal{B}}{dz}\boldsymbol{{v}}_b
+\boldsymbol{{v}}_a\sp\dagger\mathcal{B}
\frac{d\boldsymbol{{v}}_b}{dz}
=\boldsymbol{{v}}_a\sp\dagger\left(
 \frac{d\mathcal{B}}{dz} -(\mathcal{H}+\mathcal{F})\mathcal{B}-\mathcal{B}(\mathcal{H}+\mathcal{F})
  \right)\boldsymbol{{v}}_b.
\end{align*}
and thus we seek to relate the operators $\mathcal{A}$ and
$\mathcal{B}$ by
\[\mathcal{A}=
\frac{d\mathcal{B}}{dz}
-(\mathcal{H}+\mathcal{F})\mathcal{B}-\mathcal{B}(\mathcal{H}+\mathcal{F}).
\]
Introduce the operator $\mathcal{D}$ by
\[  \mathcal{D} (\mathcal{R})= \frac{d\mathcal{R}}{dz}
-(\mathcal{H}+\mathcal{F})\mathcal{R}-\mathcal{R}(\mathcal{H}+\mathcal{F}).\]
We shall use the following relations
\begin{equation}
\mathcal{F}^2=-1_2\otimes
\sum_{i=1}^3T_iT_i-\im\sum_{i,j,k=1}^3\epsilon_{ijk}\,\sigma_k\otimes
T_iT_j\label{rel1}
\end{equation}
and
\begin{equation}
\frac{d\mathcal{F}}{dz}=\im\sum_{i,j,k=1}^3\epsilon_{ijk}\, \sigma_k\otimes
T_iT_j\label{rel2}
\end{equation}
Therefore
\[  \mathcal{F}^2+\frac{d\mathcal{F}}{dz}=-1_2\otimes\sum_{i,j=1}^3 T_iT_j \]
and
\[   \left[\mathcal{F}^2+\frac{d\mathcal{F}}{dz},\mathcal{H}\right]=0. \]

\begin{proposition} \label{panagopo1} Let
\[  \mathcal{Q}=\frac{1}{r^2} \mathcal{H}\mathcal{F}\mathcal{H}-\mathcal{F} \]
Then we have the antiderivative
\begin{equation}
\int dz\, \boldsymbol{{v}}_a\sp\dagger \boldsymbol{{v}}_b
= \boldsymbol{{v}}_a\sp\dagger \mathcal{Q}^{-1}
\boldsymbol{{v}}_b.
\end{equation}
\end{proposition}
\begin{proof}
In this case $\mathcal{A}=1_{2n}$ and we must show that
\begin{align}
\frac{d\mathcal{Q}^{-1}}{dz}-(\mathcal{H}+\mathcal{F})\mathcal{Q}^{-1}
-\mathcal{Q}^{-1}(\mathcal{H}+\mathcal{F}) =1_{2n} \label{panrel}
\end{align}
The left-hand side  of (\ref{panrel}) may be rewritten as follows
\begin{align*}
\mathcal{Q}^{-1} &\left[ -\frac{d}{dz}\left(
\frac{1}{r^2}\mathcal{H}\mathcal{F}\mathcal{H}-\mathcal{F}
\right)-(\mathcal{H}+\mathcal{F})\left(
\frac{1}{r^2}\mathcal{H}\mathcal{F}\mathcal{H}-\mathcal{F} \right)
-\left(
\frac{1}{r^2}\mathcal{H}\mathcal{F}\mathcal{H}-\mathcal{F}
\right)(\mathcal{H}+\mathcal{F}) \right]\mathcal{Q}^{-1}\\
=&\mathcal{Q}^{-1}
\left[-\frac{1}{r^2}\mathcal{H}\frac{d\mathcal{F}}{dz}
\mathcal{H}+\frac{d\mathcal{F}}{dz}
-\frac{1}{r^2}(\mathcal{H}+\mathcal{F})\mathcal{H}\mathcal{F}\mathcal{H}
-\frac{1}{r^2}\mathcal{H}\mathcal{F}\mathcal{H}(\mathcal{H}+\mathcal{F})\right.\\
&\qquad \qquad
+(\mathcal{H}+\mathcal{F})\mathcal{F}+\mathcal{F}(\mathcal{H}+\mathcal{F})\bigg]\mathcal{Q}^{-1}\\
=&\mathcal{Q}^{-1}
\left[\frac{1}{r^2}\mathcal{H}\left(\mathcal{F}^2+1_2\otimes\sum_{i,j=1}^3
T_iT_j\right)\mathcal{H}-\left(\mathcal{F}^2+1_2\otimes\sum_{i,j=1}^3T_iT_j\right)\right.\\
&\left.\qquad\ -\frac{1}{r^2}\mathcal{H}^2\mathcal{F}\mathcal{H}-\frac{1}{r^2}
\mathcal{F}\mathcal{H}\mathcal{F}\mathcal{H}+\mathcal{H}\mathcal{F}+\mathcal{F}^2
-\frac{1}{r^2}\mathcal{H}\mathcal{F}\mathcal{H}^2-\frac{1}{r^2}
\mathcal{H}\mathcal{F}\mathcal{H}\mathcal{F}+\mathcal{F}\mathcal{H}+\mathcal{F}^2
\right]\mathcal{Q}^{-1}
\end{align*}
Now $\mathcal{H}^2=r^21_{2n}$ and
\begin{align*}
\mathcal{Q}^2&=\left(\frac{1}{r^2}\mathcal{H}\mathcal{F}\mathcal{H}-\mathcal{F}\right)^2
=\frac{1}{r^4}\mathcal{H}\mathcal{F}\mathcal{H}^2\mathcal{F}\mathcal{H}+\mathcal{F}^2
-\frac{1}{r^2}\mathcal{H}\mathcal{F}\mathcal{H}\mathcal{F}
-\frac{1}{r^2}\mathcal{F}\mathcal{H}\mathcal{F}\mathcal{H}\\
&=\frac{1}{r^2}\mathcal{H}\mathcal{F}^2\mathcal{H}
-\frac{1}{r^2}\mathcal{H}\mathcal{F}\mathcal{H}\mathcal{F}
-\frac{1}{r^2}\mathcal{F}\mathcal{H}\mathcal{F}\mathcal{H}+\mathcal{F}^2
\end{align*}
Performing the appropriate cancellations we obtain the necessary
result.

\end{proof}

\begin{proposition}\label{panagopo2} Let $\mathcal{Q}$ be as in the
Proposition \ref{panagopo1} and
\[  \mathcal{S}=\mathcal{Q}^{-1} \left( z+2\mathcal{H}\frac{d}{d r^2}   \right). \]
Then we have the antiderivative
\begin{equation}
\int dz\, z \boldsymbol{{v}}_a\sp\dagger \boldsymbol{{v}}_b
= \boldsymbol{{v}}_a\sp\dagger \mathcal{S}
\boldsymbol{{v}}_b.\label{parel2}
\end{equation}
\end{proposition}
\begin{proof}
Denote
\[ \mathcal{S}_1=\mathcal{Q}^{-1} z,\qquad  \mathcal{S}_2=\mathcal{Q}^{-1} 2
\mathcal{H}\frac{d}{dr^2}.
\]
Then
\begin{align*}
\mathcal{D}(\mathcal{S}_1)&=\frac{d}{dz}\left(z\mathcal{Q}^{-1}\right)-z
(\mathcal{H}+\mathcal{F})\mathcal{Q}^{-1}-z\mathcal{Q}^{-1}(\mathcal{H}
+\mathcal{F})
=z\mathcal{D}(\mathcal{Q}^{-1})+\mathcal{Q}^{-1}=z1_{2n}+\mathcal{Q}^{-1}.
\end{align*}
Further, using (\ref{panrel}),
\begin{align*}
\mathcal{D}(\mathcal{S}_2)&=\frac{d}{dz}\left(\mathcal{Q}^{-1}
2\mathcal{H}
\frac{d}{dr^2}\right)-(\mathcal{H}+\mathcal{F})\mathcal{Q}^{-1}2\mathcal{H}
\frac{d}{dr^2}
-\mathcal{Q}^{-1}\,2\mathcal{H}\,
\frac{d}{dr^2}(\mathcal{H}+\mathcal{F})\\
&=\mathcal{D}(\mathcal{Q}^{-1})\,2\mathcal{H}\frac{d}{dr^2}
\;\;+\mathcal{Q}^{-1}(\mathcal{H}+\mathcal{F})2\mathcal{H}
\frac{d}{dr^2}
-\mathcal{Q}^{-1}\,2\mathcal{H}
\frac{d}{dr^2}(\mathcal{H}+\mathcal{F})\\
&=2\mathcal{H}\frac{d}{dr^2}+\mathcal{Q}^{-1}(2r^2+2\mathcal{F}\mathcal{H})\frac{d}{dr^2}
-\mathcal{Q}^{-1}(2r^2+2\mathcal{H}\mathcal{F})\frac{d}{dr^2}
-\mathcal{Q}^{-1}2\mathcal{H}\frac{d\mathcal{H}}{dr^2}.
\end{align*}
The last term may be expressed as
\[-\mathcal{Q}^{-1}2\mathcal{H}\frac{d\mathcal{H}}{dr^2}
=-\mathcal{Q}^{-1}\frac{d\mathcal{H}^2}{dr^2}=-\mathcal{Q}^{-1}\]
whence
\begin{align*}\mathcal{D}(\mathcal{S})=
\mathcal{D}(\mathcal{S}_1+\mathcal{S}_2)=z1_{2n}
+\left(
2\mathcal{H}+2\mathcal{Q}^{-1}\mathcal{F}\mathcal{H}-2\mathcal{Q}^{-1}\mathcal{H}\mathcal{F}
\right)\frac{d}{dr^2}.
\end{align*}
Now the expression in brackets vanishes as a consequence of
\begin{align*}
2\mathcal{H}&+2\mathcal{Q}^{-1}\mathcal{F}\mathcal{H}-2\mathcal{Q}^{-1}\mathcal{H}\mathcal{F}
=2\mathcal{Q}^{-1}\left[\left( \frac{1}{r^2} \mathcal{H}\mathcal{F}\mathcal{H} -
\mathcal{F}\right)\mathcal{H}+\mathcal{F}\mathcal{H}-\mathcal{H}\mathcal{F}\right]\\
&=2\mathcal{Q}^{-1}\left[\mathcal{H}\mathcal{F}-\mathcal{F}\mathcal{H}
+\mathcal{F}\mathcal{H} -\mathcal{H}\mathcal{F}  \right]=0
\end{align*}
and the result follows.
\end{proof}

\begin{proposition}\label{panagopo3} Let $\mathcal{Q}$ be as in the
Proposition \ref{panagopo1} Then the antiderivative
\begin{align}
\int \boldsymbol{{v}}_a\sp\dagger\frac{\partial}{\partial
x_i}\boldsymbol{{v}}_b dz
&=\boldsymbol{{v}}_a\sp\dagger\mathcal{Q}^{-1} \left[
\frac{\partial}{\partial x_i}+\mathcal{H}\frac{z}{r^2}\,
x_i+\mathcal{H}\frac{\imath}{r^2} \left(
\boldsymbol{x}\times\boldsymbol{\nabla}\right)_i \right]\boldsymbol{{v}}_b
.\label{parel3}
\end{align}
\end{proposition}

\begin{proof}
Let $L=L_1+L_2+L_3$ with
\[ L_1=\mathcal{Q}^{-1}\frac{\partial}{\partial x_i},\quad
L_2= \mathcal{Q}^{-1}\mathcal{H}\frac{z}{r^2}\,x_i,\quad
L_3=\mathcal{Q}^{-1}\mathcal{H}\frac{\imath}{r^2}\left( \boldsymbol{x}\times
\boldsymbol{\nabla} \right)_i
\]
We compute $\mathcal{D}(L_i)$, $i=1,2,3$. First
\begin{align*}
\mathcal{D}(L_1)&=\frac{d}{dz}\left( \mathcal{Q}^{-1} \frac{\partial
}{\partial x_i} \right)
-(\mathcal{H}+\mathcal{F})\mathcal{Q}^{-1}\frac{\partial}{\partial
x_i}-\mathcal{Q}^{-1}\frac{\partial }{\partial
x_i}(\mathcal{H}+\mathcal{F})\\
&=\left[
\frac{d\mathcal{Q}^{-1}}{dz}-(\mathcal{H}+\mathcal{F})\mathcal{Q}^{-1}
-\mathcal{Q}^{-1}(\mathcal{H}+\mathcal{F})
\right]\frac{\partial}{\partial x_i}-\mathcal{Q}^{-1}\frac{\partial
\mathcal{H}}{\partial x_i}\\
&=1_{2n}\frac{\partial}{\partial x_i}-\mathcal{Q}^{-1}\frac{\partial
\mathcal{H}}{\partial x_i}
\end{align*}
where we use Proposition \ref{panagopo1}. Next
\begin{align*}
\mathcal{D}(L_2)&=\frac{d}{dz}\left( \mathcal{Q}^{-1}
\mathcal{H}\frac{z}{r^2}x_i \right)-(\mathcal{H}+\mathcal{F})
\mathcal{Q}^{-1} \mathcal{H}\frac{z}{r^2}x_i-\mathcal{Q}^{-1}
\mathcal{H}\frac{z}{r^2}x_i(\mathcal{H}+\mathcal{F})\\
&=\left[
\frac{d\mathcal{Q}^{-1}}{dz}-(\mathcal{H}+\mathcal{F})\mathcal{Q}^{-1}
-\mathcal{Q}^{-1}(\mathcal{H}+\mathcal{F})
\right]\mathcal{H}\frac{z}{r^2}x_i\\
&\qquad +\mathcal{Q}^{-1}(\mathcal{H}+\mathcal{F})\mathcal{H}\frac{z}{r^2}x_i
-\mathcal{Q}^{-1}\mathcal{H}\frac{z}{r^2}x_i(\mathcal{H}+\mathcal{F})
+\mathcal{Q}^{-1}\mathcal{H}\frac{x_i}{r^2}\\
&=\mathcal{H}\frac{z}{r^2}x_i+\mathcal{Q}^{-1}(r^2+\mathcal{F}\mathcal{H})\frac{z}{r^2}x_i
-\mathcal{Q}^{-1}(r^2+\mathcal{H}\mathcal{F})\frac{z}{r^2}x_i+\mathcal{Q}^{-1}\mathcal{H}\frac{x_i}{r^2}\\
&=\mathcal{Q}^{-1}\mathcal{H}\frac{x_i}{r^2}+\frac{z}{r^2}x_i(
\mathcal{H}+\mathcal{Q}^{-1}\mathcal{F}\mathcal{H}-\mathcal{Q}^{-1}\mathcal{H}\mathcal{F})
\end{align*}
Now the expression in brackets vanishes (see the proof of
Proposition \ref{panagopo2}). Therefore
\[\mathcal{D}(L_2)=\mathcal{Q}^{-1}\mathcal{H}\frac{x_i}{r^2}.\]
Next we calculate $\mathcal{D}(T_3)$ for $i=1$. We have
\begin{align*}
\mathcal{D}(L_3)&=\frac{d}{dz}\left[ \mathcal{Q}^{-1}
\mathcal{H}\frac{\imath}{r^2}\left( x_2\frac{\partial}{\partial
x_3}-x_3\frac{\partial}{\partial x_2} \right) \right]\\
&\quad -(\mathcal{H}+\mathcal{F})\left[ \mathcal{Q}^{-1}
\mathcal{H}\frac{\imath}{r^2}\left( x_2\frac{\partial}{\partial
x_3}-x_3\frac{\partial}{\partial x_2} \right) \right]
-\left[ \mathcal{Q}^{-1} \mathcal{H}\frac{\imath}{r^2}\left(
x_2\frac{\partial}{\partial x_3}-x_3\frac{\partial}{\partial x_2}
\right) \right](\mathcal{H}+\mathcal{F})\\
&=\left[  \frac{d\mathcal{Q}^{-1}}{dz}
-(\mathcal{H}+\mathcal{F})\mathcal{Q}^{-1}
-\mathcal{Q}^{-1}(\mathcal{H}+\mathcal{F})\right]\mathcal{H}\frac{\imath}{r^2}\left(
x_2\frac{\partial}{\partial x_3}-x_3\frac{\partial}{\partial x_2}
\right)\\
&\quad +\mathcal{Q}^{-1}(\mathcal{H}+\mathcal{F})\mathcal{H}\frac{\imath}{r^2}\left(
x_2\frac{\partial}{\partial x_3}-x_3\frac{\partial}{\partial x_2}
\right)
-\mathcal{Q}^{-1}\mathcal{H}\frac{\imath}{r^2}\left(
x_2\frac{\partial}{\partial x_3}-x_3\frac{\partial}{\partial x_2}
\right)(\mathcal{H}+\mathcal{F})\\
&=\mathcal{H}\frac{\imath}{r^2}\left( x_2\frac{\partial}{\partial
x_3}-x_3\frac{\partial}{\partial x_2}
\right)+\mathcal{Q}^{-1}(\mathcal{H}+\mathcal{F})\mathcal{H}\frac{\imath}{r^2}\left(
x_2\frac{\partial}{\partial x_3}-x_3\frac{\partial}{\partial x_2}
\right)\\
&\qquad-\mathcal{Q}^{-1}\mathcal{H}(\mathcal{H}+\mathcal{F})\frac{\imath}{r^2}\left(
x_2\frac{\partial}{\partial x_3}-x_3\frac{\partial}{\partial x_2}
\right)-\mathcal{Q}^{-1}\mathcal{H}\frac{\imath}{r^2}\left(
x_2\frac{\partial}{\partial x_3}-x_3\frac{\partial}{\partial x_2}
\right)\mathcal{H}\\
&=\left[\mathcal{H}+\mathcal{Q}^{-1}(\mathcal{H}^2+\mathcal{F}\mathcal{H})
-\mathcal{Q}^{-1}(\mathcal{H}^2+\mathcal{H}\mathcal{F})\right]
\frac{\imath}{r^2}\left( x_2\frac{\partial}{\partial
x_3}-x_3\frac{\partial}{\partial x_2} \right)\\
&\quad-\mathcal{Q}^{-1}\mathcal{H}\frac{\imath}{r^2}\left(
x_2\frac{\partial}{\partial x_3}-x_3\frac{\partial}{\partial x_2}
\right)\mathcal{H}.
\end{align*}
Finally we obtain
\[  \mathcal{D}(L_3)= -\mathcal{Q}^{-1}\mathcal{H}\frac{\imath}{r^2}\left(
x_2\frac{\partial}{\partial x_3}-x_3\frac{\partial}{\partial x_2}
\right)\mathcal{H}. \] Altogether we have
\begin{align*}
\mathcal{D}(T)=1_{2n}\frac{\partial}{\partial
x_1}+\mathcal{Q}^{-1}\left\{-\frac{\partial \mathcal{H}}{\partial
x_1}+\mathcal{H}\frac{x_1}{r^2}-\mathcal{H}\frac{\imath}{r^2} \left(
x_2\frac{\partial}{\partial x_3}-x_3\frac{\partial}{\partial x_2}
\right)\mathcal{H}\right\}
\end{align*}
Multiplying the expression in the parentheses by $-r^2$ gives
\begin{align*}
-\left\{\cdot\right\}r^2&=-\sigma_1\otimes 1_2(x_1^2+x_2^2+x_3^2)
+\sigma_1\otimes 1_2x_1^2+\sigma_2\otimes 1_2x_1
x_2+\sigma_3\otimes 1_2x_1 x_3\\
&\quad
-\imath(\sigma_1\otimes 1_2 x_1 +\sigma_2\otimes 1_2 x_2+\sigma_3\otimes 1_2 x_3  )
\times (x_2\sigma_3\otimes 1_2-x_3\sigma_2\otimes 1_2)
\end{align*}
which vanishes by standard relations, proving the result.

\end{proof}


\begin{thebibliography}{BPP82}

\bibitem[BE06]{bren06}
H.~W. Braden and V.~Enolski, \emph{Remarks on the complex geometry of
  3-monopole}, 1--65, arXiv: math-ph/0601040.

\bibitem[Dub77]{dubrovin77}
B.~A. Dubrovin, \emph{Completely integrable systems related to matrix operators
  and abelian varieties}, Funk. Anal. Appl. \textbf{11} (1977), no.~4, 28--41.

\bibitem[ES89]{ersi89}
N.~Ercolani and A.~Sinha, \emph{Monopoles and {B}aker functions},
  Commun.Math.Phys. \textbf{125} (1989), 385--416.

\bibitem[Hit82]{hitchin82}
N.~J. Hitchin, \emph{Monopoles and {G}eodesics}, Commun.Math.Phys. \textbf{83}
  (1982), 579--602.

\bibitem[Hit83]{hitchin83}
\bysame, \emph{On the {C}onstruction of {M}onopoles}, Commun.Math.Phys.
  \textbf{89} (1983), 145--190.

\bibitem[Nah82]{nahm82}
W.~Nahm, \emph{The construction of all self-dual multimonopoles by the {ADHM}
  method}, World Scientific, Singapore, 1982.

\bibitem[Pan83]{panagopo83}
H.~Panagopoulos, \emph{Multimonopoles in arbitrary gauge groups and the
  complete $su(2)$ two-monopole system}, Phys.Rev.D \textbf{28} (1983), no.~2,
  380--384.

\bibitem[WY06]{weinbyi06}
Eric~J. Weinberg and Piljin Yi, \emph{Magnetic {M}onopole {D}ynamics,
  {S}upersymmetry and {D}uality}, 1--251, arXiv: hep-th/0609055.

\end{thebibliography}

\providecommand{\bysame}{\leavevmode\hbox
to3em{\hrulefill}\thinspace}
\providecommand{\MR}{\relax\ifhmode\unskip\space\fi MR }
\providecommand{\MRhref}[2]{%
  \href{http://www.ams.org/mathscinet-getitem?mr=#1}{#2}
}
\providecommand{\href}[2]{#2}

\end{document}